\documentclass[aps,prl,twocolumn,showpacs]{revtex4-1} 
\usepackage{graphicx,amsmath}

\begin{document}


\title{Mechanism for the large conductance modulation in electrolyte gated thin gold films} 



\author{Trevor Petach}
\affiliation{Department of Physics, Stanford University, Palo Alto, CA 94305, USA}
\author{Menyoung Lee}
\affiliation{Department of Physics, Stanford University, Palo Alto, CA 94305, USA}
\author{Ryan Davis}
\affiliation{SLAC National Accelerator Laboratory, Menlo Park, CA 94205, USA}
\author{Apurva Mehta}
\affiliation{SLAC National Accelerator Laboratory, Menlo Park, CA 94205, USA}
\author{David Goldhaber-Gordon}
\affiliation{Department of Physics, Stanford University, Palo Alto, CA 94305, USA}



\date{\today}

\begin{abstract}
Electrolyte gating using ionic liquid electrolytes has recently generated considerable interest as a method to achieve large carrier density modulations in a variety of materials. In noble metal thin films, electrolyte gating results in large changes in sheet resistance. The widely accepted mechanism for these changes is the formation of an electric double layer with a charged layer of ions in the liquid and accumulation or depletion of carriers in the thin film. We report here a different mechanism. In particular, we show using x-ray absorption near edge structure (XANES) that the previously reported large conductance modulation in gold films is due to reversible oxidation and reduction of the surface rather than the charging of an electric double layer. We show that the double layer capacitance accounts for less than 10\% of the observed change in transport properties. These results represent a significant step towards understanding the mechanisms involved in electrolyte gating.
\end{abstract}


\maketitle 

%

Modulating carrier density using the electric field effect is a cornerstone of condensed matter physics. To achieve the largest carrier density modulations of more than 15 $\mu$C/cm$^2$ ($10^{14}$ carriers/cm$^2$), experimenters have turned to electrolyte gating with ionic liquids \cite{Sagmeister2006}. This technique involves immersing the channel material and a gate electrode in an ionic liquid to form an ``electric double layer transistor'' \cite{Prassides2011}. Since ionic liquids have wide electrochemical stability windows, large potentials can be applied between the gate and the channel without chemically altering the liquid \cite{Sato2004}. Using ionic liquid electrolyte gating, significant changes have been observed in the transport properties of organic semiconductors \cite{Cho2008a}, oxides \cite{Lee2011f, Nakano2012, Ha2013}, and even metals \cite{Sagmeister2006, Daghero2012, Nakayama2012}. 

The widely accepted mechanism for the observed changes in transport in these electrolyte gating experiments is the charging of the electric double layer at the channel surface \cite{Prassides2011}. In this mechanism, when a voltage is applied between the gate and the channel, a charged layer of ions forms on the channel surface and a corresponding change in carrier density occurs in the channel. The charged layer of ions and the net charge in the channel are known as the electric double double layer, and the double layer capacitance gives the charge density that would be accumulated in each layer per unit potential. However, the reported carrier densities of more than 150 $\mu$C/cm$^2$ (10$^{15}$ carriers/cm$^2$) at moderate potential ($<5$ V) exceed typical values of the double layer capacitance \cite{Bard1980} by more than an order of magnitude, suggesting other mechanisms may be involved.

Recently, a careful study of electrolyte gated vanadium dioxide suggested that migration of oxygen vacancies, rather than the charging of the electric double layer, is the dominant mechanism for the observed metal to insulator transition in this system \cite{Jeong2013}. However, this mechanism can not explain the large electrolyte gating effect in metals.

In this Letter, we show using x-ray absorption near edge structure (XANES) that reversible oxidation of a thin surface layer fully explains the dramatic changes in transport properties of electrolyte gated gold films. We use electrochemical impedance spectroscopy to show that charging of the electric double layer explains less than 10\% of the observed change in carrier density.

\begin{figure}
\includegraphics[width=3.375in]{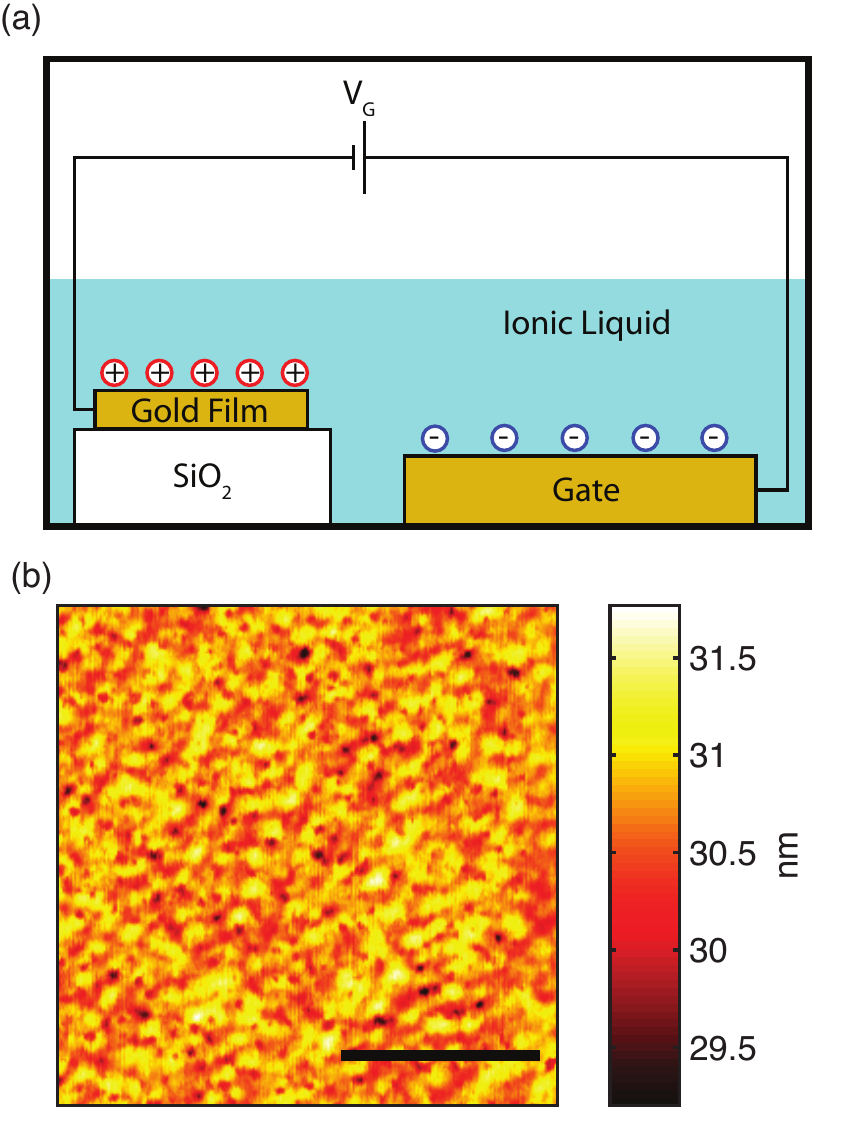}
\caption{(a) Device schematic. A gate electrode and thin gold film are immersed in an ionic liquid inside a controlled atmosphere (high vacuum for transport measurements and high purity helium gas for XANES measurements). Gold wires are used to make contacts. (b) An atomic force micrograph of the thin gold film. Scale bar is 200 nm.}
\label{fig:device}
\end{figure}

We immerse a thin gold film (channel) in an ionic liquid and use a larger gold electrode (gate) to apply a gate voltage. As shown in Fig. \ref{fig:device}(a), this system has 2 electrodes. The channel serves as the working electrode, and the gate serves as the counter electrode. We study the properties of the channel (carrier density, sheet resistance, and oxidation state).

Gold films are lithographically defined with standard e-beam techniques and lift off. They are deposited on SiO$_2$/Si (Silicon Quest International, 300 nm wet thermal oxide) in an e-beam evaporator (custom Kurt J. Lesker Co., 1.5 \AA/s). Films for transport are 30 nm thick, and films for XANES are 100 nm thick. The root-mean-square surface roughness is less than 1 nm as measured by atomic force microscopy (see Fig. \ref{fig:device}(b)). The films are cleaned with acetone, isopropanol, de-ionized water, and 60 seconds in an oxygen plasma asher (March PX250, direct plasma, 300 W, 120 mbar O$_2$). Leads for transport measurements are capped with a hard baked (300 C for 30 mins) S1813 photoresist mask which leaves only the channel and the gate electrode in contact with the liquid. Leads for XANES and EIS measurements do not have photoresist masks, and the gate and channel areas are much larger.

The ionic liquid used in all experiments is N,N-diethyl-N-methyl-N-2-methoxyethyl ammonium tetrafluoroborate (DEME-BF4, Kanto Chemical Company). The liquid is dried in a vacuum oven (120 C, 100 Torr) overnight before being transferred onto the gold films. Cyclic voltammetry (CV) measurements show no evidence of contamination (see Supplemental Material \cite{SI}).

\begin{figure}
\includegraphics[width=3.375in]{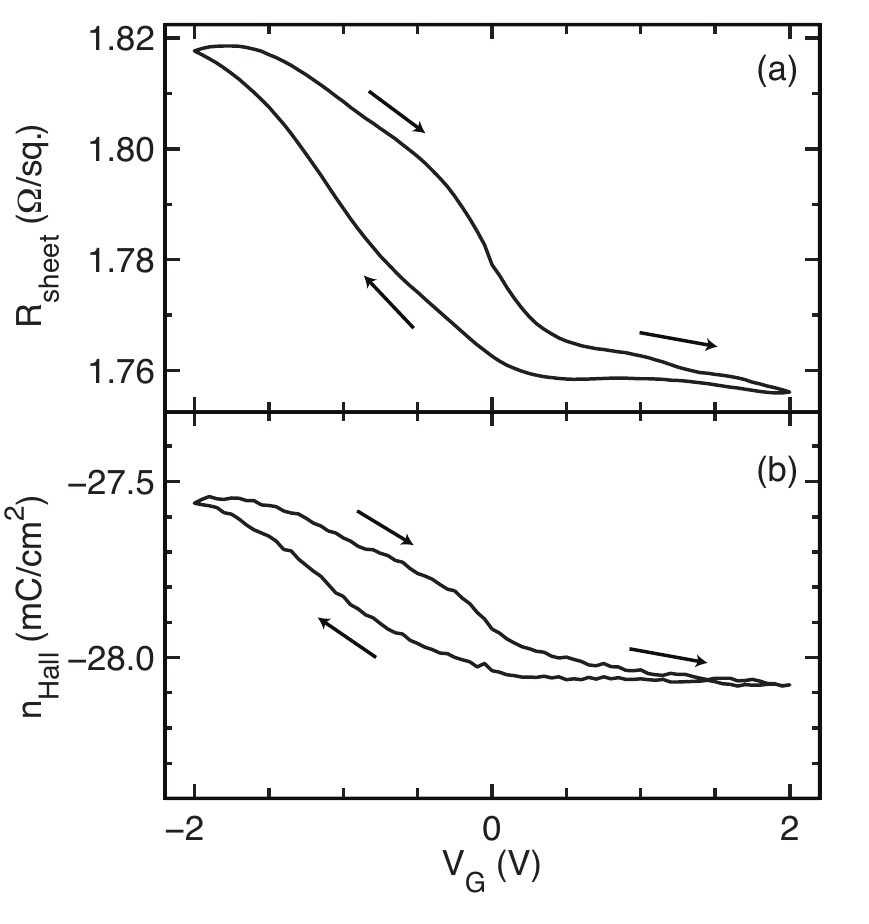}
\caption{(a) Sheet resistance and (b) Hall coefficient versus gate voltage for a 30 nm film in DEME-BF4. The charge carriers are electron-like. Sweep rate is 100 mV/s. Each curve is the average of three cycles.}
\label{fig:CV}
\end{figure}

We measure Hall coefficient and sheet resistance at room temperature in a Quantum Design PPMS using a 4 terminal lock-in technique. The reported Hall coefficient is the average of the value at $+9$ T and $-9$ T. We observe large changes in the Hall coefficient and the sheet resistance as a function of gate voltage (see Fig. \ref{fig:CV}). The largest changes occur at negative voltages. The difference in Hall coefficient between $-2$ V and 2 V suggests that we can modulate (primarily deplete) the carrier density by 510 $\mu$C/cm$^2$ ($3.2\times10^{15}$ carriers/cm$^2$ or complete depletion of 2.3 monolayers). The increase in sheet resistance is consistent with the decrease in Hall density at negative voltages. There is also a modest decrease in mobility at negative gate voltages, which we attribute to changes in surface scattering \cite{Sondheimer2001} (see Supplemental Material \cite{SI} for a discussion of mobility). These changes are comparable to the largest reported electroresistance and carrier density modulation using electrolyte gating \cite{Prassides2011} and to previous work gating noble metal channels \cite{Daghero2012}.

\begin{figure}
\includegraphics[width=3.375in]{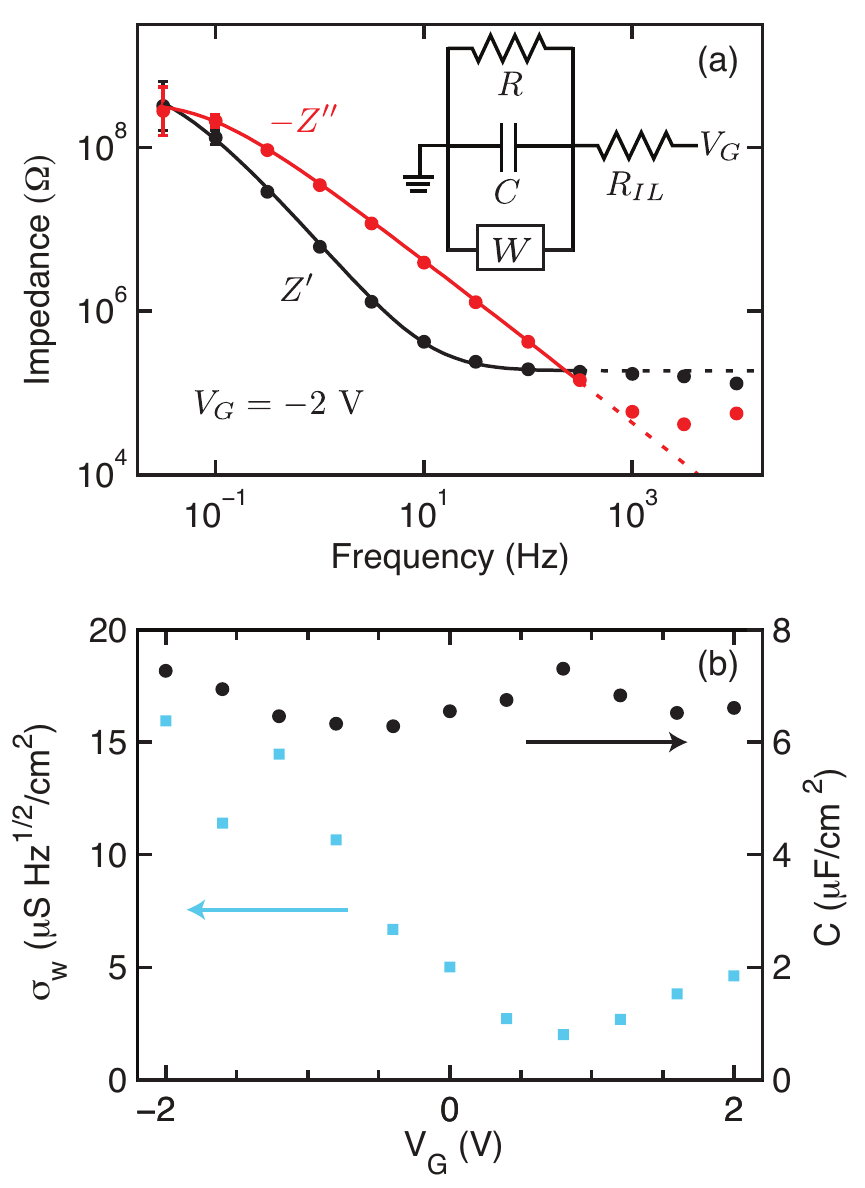}
\caption{(a) Typical electrochemical impedance spectroscopy data at $V_G = -2$ V. Error bars show 1$\sigma$ confidence intervals. Solid lines show the best fit to the model shown in the inset in the range $10^{-1.5}$ - $10^{2.5}$ Hz. Dashed lines extrapolate the model to higher frequencies. (b) Best fit differential capacitance (black circles) and Warburg conductivity (blue squares) as a function of DC gate bias. See Supplemental Material \cite{SI} for $R$ and $R_{IL}$.}
\label{fig:EIS}
\end{figure}

To investigate whether the observed changes in transport can be explained by electrostatic accumulation of carriers due to charging of the electric double layer, we use electrochemical impedance spectroscopy. Electrochemical impedance spectroscopy is a standard method to measure electric double layer capacitance \cite{Bard1980}. We use a Stanford Research System model 830 lock-in amplifier and a Yokogawa GS200 voltage source to measure admittance as a function of frequency and DC gate voltage in a Desert Cryogenics probe station at 10$^{-4}$ Torr and 295 K. A 400 mV rms sinusoid is sourced from the amplifier's internal reference, and current is measured by monitoring the voltage across a small (1 k$\Omega$) series resistor to ground. We use a 2 electrode setup. The working electrode is 100 $\times$ 500 $\mu$m$^2$, and the counter electrode is much larger ($\approx 10 \times$) than the working electrode. This size difference allows us to neglect the impedance of the larger electrode. The gate voltage $V_G$ is the potential difference between the counter and the working electrode (the sign of gate voltage is considered positive when the counter electrode is at a more positive voltage than the working electrode). 

We measure impedance as a function of frequency from 32 mHz to 10 kHz and report the in phase ($Z^{\prime}$) and out of phase ($Z^{\prime\prime}$) components as a function of frequency. A typical measurement is shown in Fig. \ref{fig:EIS}(a). We model the ionic liquid - gold interface as a resistor ($R$), a capacitor ($C$), and a Warburg impedance ($Z_W = W/\sqrt{f} - iW/\sqrt{f})$) in parallel. The resistor models a leakage current. The capacitor models the charge accumulated in the electric double layer, and the Warburg impedance models charge flowing through the interface due to impurities diffusing to the surface and then being oxidized or reduced. The finite ionic conductivity of the liquid and the contact resistance are modeled as a single resistor ($R_{IL}$) in series with the interface (see Fig. \ref{fig:EIS}(a), inset). We fit the data in the range from 32 mHz to 320 Hz using a least squares regression algorithm with each point weighted inversely to the measurement uncertainty. We report $R$ and $W$ as conductivities $\sigma$ and $\sigma_W$, respectively. We normalize $\sigma$, $\sigma_W$, and $C$ to the area of the working electrode. We include a discussion of the Warburg conductivity and report $R$ and $R_{IL}$ in the Supplemental Material \cite{SI}. 

The best fit double layer capacitance and Warburg conductivity are shown in Fig. \ref{fig:EIS}(b). The double layer capacitance ranges from $6 - 8$ $\mu$F/cm$^2$ and is roughly constant as a function of gate voltage. The slight rise in capacitance and dip in Warburg conductivity at $V_G = 0.8$ V may mark the potential of zero charge \cite{Kornyshev2007, Bazant2011}. The model fits well up to 1 kHz, suggesting that the double layer capacitance is constant up to 1 kHz. 

From our capacitance fit, the change in carrier density from 0 V to $-2$ V due to charging of the electric double layer is 14 $\mu$C/cm$^2$ ($9\times10^{13}$ carriers/cm$^2$), 35 times smaller than the observed change in Hall coefficient. We conclude that the observed changes in Hall coefficient can not be explained by charging of the electric double layer.

However, the time scales in many electrolyte gating experiments can be as long as minutes \cite{Nakano2012} or hours \cite{Daghero2012}. At these long time scales, Faradaic processes (involving reduction or oxidation of a species at the electrode, which we have modeled using a Warburg term) can dominate the total current flowing across the interface. In our case, at $-2$ V, Faradaic processes dominate the total interfacial current at frequencies slower than 4.8 Hz. If the Faradaic processes chemically modify the channel surface, then for slowly varying gate voltages, this chemical modification may have a much larger effect on the transport properties than the charging of the electric double layer.

\begin{figure}
\includegraphics[width=3.375in]{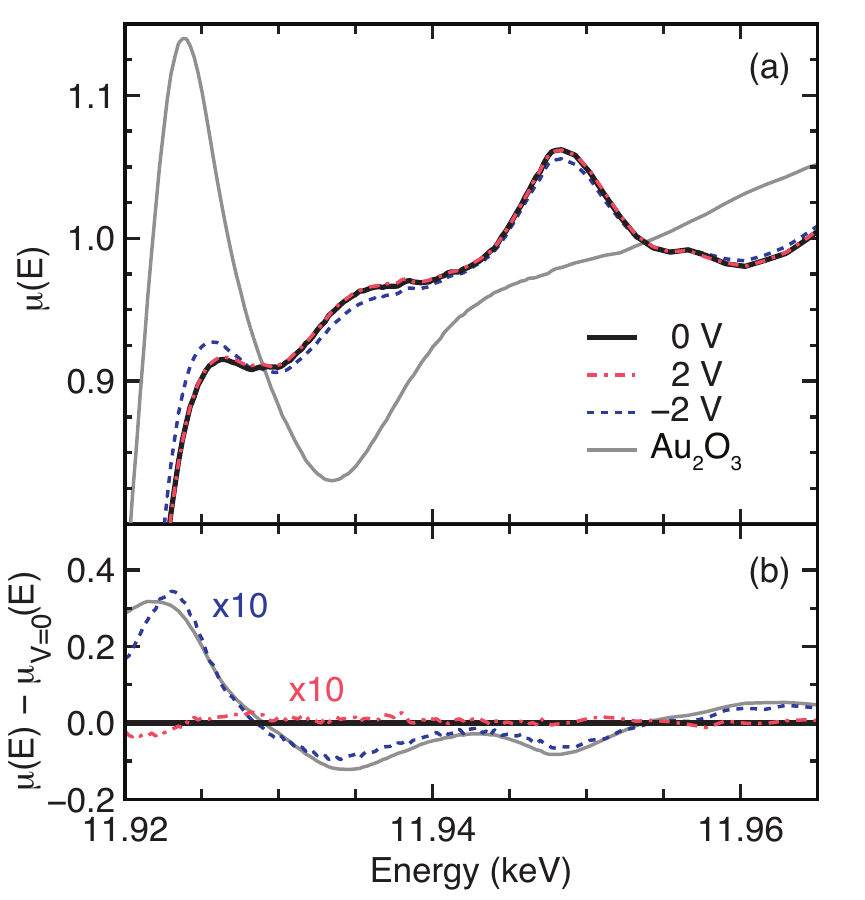}
\caption{Formation of Au$_2$O$_3$ surface layer at negative gate voltage.  (a) Normalized absorption coefficient versus incident x-ray energy (XANES) and (b) difference between 0 V spectrum and other spectra, 0.08$^{\circ}$ grazing angle. Au$_2$O$_3$ spectrum from \cite{Min2014}.}
\label{fig:XAS}
\end{figure}

Using grazing incidence XANES, we observe the chemical state of the surface gold atoms \emph{in situ}. Near an absorption edge, the absorption coefficient $\mu$ changes as a function energy in a way that depends on the electronic structure of the atom, including its oxidation state and local chemical environment \cite{Koningsberger1988}. By comparing the observed spectrum to a set of known spectra, the relative abundance of different compounds can be inferred. XANES is collected at beamline 11-2 at the Stanford Synchrotron Light Source (SSRL) in a grazing incidence geometry using a 100 element liquid nitrogen cooled Ge detector. The sample chamber is purged with 0.5 standard cubic feet per hour of 99.99\% helium gas. The data are normalized using the SixPack software package.

The gold L3 edge grazing incidence (0.08$^{\circ}$) XANES spectrum at several gate voltages is shown in Fig. \ref{fig:XAS}(a). At negative gate voltages, we observe a rise in the white line intensity (the feature at 11.925 keV) and small changes in other peak intensities. These changes are reversible and do not change during repeated scans (20 mins each) at a fixed gate voltage (see Supplemental Material \cite{SI}). The increase in white line intensity indicates that the d orbitals of some gold atoms have been depleted from their metallic d$^{10}$ configuration \cite{Friebel2011}. The changes in other peak intensities indicate the presence of atoms besides gold in the first coordination shell \cite{Salama1996, Sharma2007, Weiher2003}. Taken together, these observations suggest electron transfer from gold to another atom and formation of a bond. We consider reference spectra from gold-halogen compounds \cite{Salama1996}, gold-sulfur compounds \cite{Sharma2007}, and gold-oxygen compounds \cite{Weiher2003}. Of these, only the gold-oxygen compounds can explain the general features of the $-2$ V spectrum, and Au$_2$O$_3$ is the best match.

For quantitative analysis, we measure a Au$_2$O$_3$ reference sample prepared by pulsed oxidation of gold foil in 0.5 M sulfuric acid. The potential is stepped from 0.7 V to 2.7 V (vs. Hg/Hg$_2$SO$_4$) at 1 kHz for 1 h. Such a recipe produces primarily Au$_2$O$_3$ instead of Au$_2$O \cite{Cadle1972}. Au$_2$O$_3$ is known to be unstable in ambient conditions \cite{Tsai2003}, so the sample is measured immediately after preparation. As shown in Fig. \ref{fig:XAS}(b), the difference between the $-2$ V and 0 V spectra closely matches the difference between the Au$_2$O$_3$ and 0 V spectra, indicating the top layer of gold oxidizes to Au$_2$O$_3$ at negative voltages. Presuming that the observed $-2$ V spectrum is a linear combination of Au$_2$O$_3$ and metallic gold (0 V), we perform a least squares regression to determine the amount of Au$_2$O$_3$ present. The best fit is 91.5\% metallic gold spectrum and 8.5\% Au$_2$O$_3$ spectrum. 

To estimate the number of oxidized gold atoms at $-2$ V, we model the film as a slab of Au$_2$O$_3$ sitting on a slab of metallic gold. With atomic force microscopy we measure surface roughness with a height distribution of 0.7 nm full width at half maximum and a lateral correlation length of 26 nm full width at half maximum (see Fig. \ref{fig:device}(b)). This roughness suggests that the typical deviation of the surface normal from the average is 1.5$^{\circ}$, much greater than the critical angle of 0.2$^{\circ}$ \cite{Henke1993}. Thus, we use a rough surface model \cite{VonBohlen2009} to calculate the penetration depth perpendicular to the surface, which we find to be 4.2 nm (using the attenuation length of bulk gold at 11.95 keV). Based on our linear combination fit and this penetration depth, we estimate that the Au$_2$O$_3$ slab is 3.6 $\pm 1.8$ \AA\ thick (1.6 monolayers). The largest uncertainty in this estimate arises because the Au$_2$O$_3$ standard has a more porous surface morphology, and thus less self absorption (see Supplemental Material \cite{SI}) than the gold film. As expected, the Au$_2$O$_3$ is confined to the surface: at larger grazing angle (3.0$^{\circ}$, penetration depth 150 nm), we probe the full sample depth of 100 nm instead of the top 4.2 nm, and we observe no change in the XANES spectrum at any gate voltage (see Supplemental Material \cite{SI}).

Au$_2$O$_3$ is known to be a semiconductor with a band gap of $\sim$0.5 eV \cite{Koslowski2001}. Thus, at room temperature, the intrinsic carrier density in Au$_2$O$_3$ is negligible compared to the carrier density in metallic gold. Since each metallic gold atom contributes one carrier, oxidizing one gold atom removes one carrier. Using the bulk carrier density of gold ($5.9\times10^{22}$ carriers/cm$^3$ \cite{Ashcroft1976}) and the measured oxide thickness, we estimate that 360 $\pm 180 \mu$C/cm$^2$ ($2.2\times 10^{15}$ carriers/cm$^2$) are removed from the film due to oxidation.

From these XANES measurements, we conclude that the top 2 monolayers of metallic gold oxidize to Au$_2$O$_3$ at $-2$ V. This observation is consistent with the high Warburg conductivity at negative gate voltages, which suggests some sort of chemical reaction. This observation also explains the large change in Hall coefficient: the number of carriers depleted due to oxidation matches (within the uncertainty) the change in Hall coefficient. We emphasize that while oxidizing the gold surface does change the carrier density, it does not imply a large charge density in the electric double layer.

In conclusion, we have shown that surface oxidation and reduction is the dominant mechanism for large changes in transport in electrolyte gated gold films. Charging of the electric double layer has a much smaller effect on transport than the oxidation of the surface. We show that the double layer capacitance is $6-8$ $\mu$F/cm$^2$, 35 times too small to explain the observed changes in transport. 

Since there is a discrepancy between the observed double layer capacitance and the observed changes in transport in many electrolyte gating experiments, it is likely that surface oxidation plays an important role in many electrolyte gating experiments. Electrochemical impedance spectroscopy shows that the double layer capacitance is constant up to 1 kHz, suggesting that one possible way to distinguish electrostatic effects from electrochemical effects is to measure transport as a function of gate frequency.


%

\vspace{10pt}
\begin{acknowledgments}
This work was supported as part of CNEEC, an EFRC funded by the U.S. Department of Energy, BES under Award DE-SC0001060. T. P. acknowledges fellowship support from the Department of Defense through the NDSEG Program. We thank Xiaoquan Min, Christina Li, and Matt Kannan for providing the Au$_2$O$_3$ reference spectrum.
\end{acknowledgments}


%

\end{document}